\documentclass[12pt]{iopart}
\usepackage{graphicx}
\usepackage{setstack,cite}
\usepackage{iopams}
\begin{document}
\newcommand{\n}{\noindent}
\newcommand{\nn}{\nonumber}
\newcommand{\p}{\partial}
\newcommand{\bea}{\begin{eqnarray}}
\newcommand{\eea}{\end{eqnarray}}
\newcommand{\gp}{Gladwin Pradeep R}
\title[Nonlocal symmetries]
{Nonlocal symmetries of a class of scalar and coupled nonlinear ordinary differential equations of any order}
\author{  R. Gladwin Pradeep, V. K. Chandrasekar, M. Senthilvelan
 and M. Lakshmanan}
\ead{lakshman@cnld.bdu.ac.in}
\address{Centre for Nonlinear Dynamics, Department of Physics,
Bharathidasan University, Tiruchirappalli - 620 024, India }

\begin{abstract}
In this paper we devise a systematic procedure to obtain nonlocal symmetries of a class of scalar nonlinear ordinary differential equations (ODEs) of arbitrary order related to linear ODEs through  nonlocal relations.  The procedure makes use of the Lie point symmetries of the linear ODEs and the nonlocal connection to deduce the nonlocal symmetries of the corresponding nonlinear ODEs.  Using these nonlocal symmetries we obtain reduction transformations and reduced equations to specific examples.  We find the reduced equations can be explicitly integrated to deduce the general solutions for these cases.  We also extend this procedure to coupled higher order nonlinear ODEs with specific reference to second order nonlinear ODEs.
\end{abstract}
%\submitto{\JPA}

\maketitle
\section{Introduction}
During the past two decades or so there has been increased interest to find the nonlocal symmetries of ordinary differential equations \cite{abraham:1992,abraham,abraham-shrauner:4809,govinder:1995}.  Consider an $n^{th}$ order ordinary differential equation (ODE) 
\bea
A\equiv \frac{d^nx}{dt^n}+F(t,x,x^{(1)},x^{(2)},\ldots,x^{(n-1)})=0,\qquad x^{(k)}=\frac{dx^k}{dt^k},
\eea
  to be invariant under the infinitesimal transformations $X=x+\epsilon\eta(t,x)$, $T=t+\epsilon\xi(t,x)$, where $\xi$ and $\eta$ are the infinitesimal point symmetries associated with the given equations.  The vector field associated with the Lie point symmetry \cite{hydon,ibragimov:1999} is then $V=\xi(t,x)\frac{\partial}{\partial t}+\eta(t,x)\frac{\partial}{\partial x}$.

The Lie point symmetries $\xi(t,x)$ and $\eta(t,x)$ are obtained by solving the invariant condition, that is
\begin{eqnarray}
V^{(n)}(A)|_{A=0}=0,\label{invariance}
\end{eqnarray}
where
\begin{numparts}
\begin{eqnarray}
&&V^{(n)}=\xi\frac{\partial}{\partial t}+\eta\frac{\partial}{\partial x}+\eta^{(1)}\frac{\partial}{\partial x^{(1)}}+\ldots+\eta^{(n)}\frac{\partial}{\partial x^{(n)}},\\
&& \eta^{(k)}=\frac{d\eta^{(k-1)}}{dt}-x^{(k)}\frac{d\xi}{dt},\,\,\,\eta^{(0)}=\eta,
\end{eqnarray}
\end{numparts}
is the $n^{th}$ prolongation.  Thus the point symmetries $\xi(t,x)$ and $\eta(t,x)$ can be calculated in an algorithmic way for a differential equation of any order.  However, there exist more generalized symmetries such as contact symmetries, involving derivatives of $x$ in $\eta$ and $\xi$, and nonlocal symmetries, involving nonlocal terms in $\eta$ and $\xi$.

The vector field of the nonlocal symmetries is of the form
$V=\xi(t,x,\int u(t,x) dt)\frac{\partial}{\partial t}+\eta(t,x,\int u(t,x) dt)\frac{\partial}{\partial x}$.   Unlike the case of point symmetries these nonlocal symmetries cannot be determined completely in an algorithmic way because of the presence of nonlocal terms.  
The role of such nonlocal symmetries in the integration of differential equations was illustrated by Abraham-Shrauner et al. and later on by others \cite{abraham:1992,abraham,abraham-shrauner:4809,govinder:1995}.   Conventionally such nonlocal symmetries are explored either by reducing or increasing the order of the equation \cite{abraham-shrauner:4809,abraham:1992}.  
Methods to identify nonlocal symmetries of partial differential equations were also developed alongside \cite{cieslinski:1992,krasilshchik1,krasilshchik2}.
 In a recent paper, the nonlocal symmetries of two higher dimensional generalizations of the modified Emden equations were studied  \cite{karasu}. The first system is made up of two uncoupled modified Emden equations. The second system is obtained by assuming the variable of the scalar modified Emden equation to be complex and separating the real and imaginary parts \cite{ali:2009}.  

In this paper we devise a procedure to identify the nonlocal symmetries of a class of ODEs which includes the Riccati and Abel chains \cite{carinena:2009}.   In this procedure we nonlocally map the symmetries of the given $n^{th}$ order nonlinear ODE to the point symmetries of the associated $n^{th}$ order linear ODE, thereby preserving the order of the equation.
We also show with the aid of specific examples (second order, third order and coupled second order ODEs), one can obtain the known general solution of a given equation  using the associated nonlocal symmetries identified by this procedure.
In developing this procedure we make a judicious use of our earlier work on the nonlocal connection between nonlinear and linear ODEs \cite{chandrasekar:06:01:jpa} to construct the nonlocal symmetries for a given nonlinear ODE.   We show that the same procedure is applicable to any order starting from 2 to arbitrary $N$. Further, we extend the procedure to deduce the nonlocal symmetries of a class of coupled second order ODEs, which includes the coupled modified Emden equation \cite{gladwin:jpa:2009}.

 The plan of the paper is as follows.  In Sec. \ref{procedure} we describe the general procedure to obtain the nonlocal symmetries associated with a class of second order nonlinear ODEs.  Using the nonlocal symmetries we deduce the general solution for two interesting equations belonging to this class of equations.  Further, we extend the procedure to a more general class of second order ODEs.  In Sec. \ref{third-order-symmetry}, we extend the applicability of the procedure to a class of third order ODEs.  In Sec. \ref{arbitrary-order}, we apply this procedure to a class of $n^{th}$ order ODEs and deduce the associated nonlocal symmetries.
In Sec. \ref{coupled-symmetry}, we extend the procedure to a class of coupled second order ODEs and obtain their nonlocal symmetries.  Further, we deduce the general solution of the coupled modified Emden type equation using its nonlocal symmetries.  In Sec. \ref{conclusion}, we summarize our results.  In the appendix we demonstrate briefly how the nonlocal symmetries identified through the developed procedure indeed satisfies the symmetry invariant condition (\ref{invariance}).

\section{Nonlocal symmetries }
\label{procedure}
Let us consider the following class of nonlinear second order ODE,
\bea
&&\hspace{-1.3cm}\ddot{x}+(n-1)\frac{\dot{x}^2}{x}+((c_{1}+2f)+\frac{1}{n}xf_x)\dot{x}+\frac{x}{n}(f^2+c_1f+c_2)=0,\,\,f_x=\frac{\partial f}{\partial x},\label{2gen}
\eea
where $\left(\dot{}=\frac{d}{dt}\right)$, which is related to the second order linear ODE,
\bea
\ddot{U}+c_1\dot{U}+c_2U=0,\quad\left(\dot{}=\frac{d}{dt}\right)\label{horm1}
\eea
through the nonlocal transformation
\bea
U=x^ne^{\int f(x)dt}.\label{trans}
\eea
Here $c_1,\,c_2$ and $n$ are real constants and $f=f(x)$ is an arbitrary given function.
Equation (\ref{2gen}) includes many physically and mathematically interesting equations such as the modified Emden equation \cite{leach:85:01,chandrasekar:emden}, Ermakov-Pinney equation \cite{pinney} and generalized Duffing-van der Pol equation.  Equation (\ref{2gen}) reduces to the Li\'enard class of equations for the parametric choice $n=1$.  Classification of the forms of $f(x)$ for this Li\'enard class of equations admitting Lie point symmetries has been carried out in Refs. \cite{pandey:082702,pandey:102701}.  We note that for arbitrary forms of $f(x)$,  Eq. (\ref{2gen}) admits only the time translation symmetry.
In addition to the Lie point symmetries admitted by Eq. (\ref{2gen}), there exists other generalized symmetries such as contact symmetries, nonlocal symmetries and so on.  In order to explore the nonlocal symmetries associated with (\ref{2gen}), we use the identity
\begin{eqnarray}            
\frac{\dot{U}}{U}=\frac{n\dot{x}}{x}+f(x),\label{gme01}
\label {gme02}
\end{eqnarray} 
which can be directly deduced from (\ref{trans}).
\subsection{General Theory}
 The above nonlocal connection between Eqs. (\ref{2gen}) and (\ref{horm1}) allows us to deduce the nonlocal symmetries of Eq. (\ref{2gen}).  To verify this we proceed as follows.
Let $\xi$ and $\eta$ be the infinitesimal point transformations, that is $U'=U+\epsilon \eta(t,U)$, $T=t+\epsilon \xi(t,U)$, associated with the linear ODE (\ref{horm1}).  Then the symmetry vector field associated with the infinitesimal transformations read as
\begin{eqnarray}            
\Lambda=\xi \frac{\partial}{\partial t}
+\eta \frac{\partial}{\partial U},
\label {sym01}
\end{eqnarray}
and the first extension is
\begin{eqnarray}            
\Lambda^{1}=\xi \frac{\partial}{\partial t}
+\eta \frac{\partial}{\partial U}+(\dot{\eta}
-\dot{U}\dot{\xi})\frac{\partial}{\partial \dot{U}}.
\label {sym02}
\end{eqnarray}

Let us designate the symmetry vector field and its first prolongation of the nonlinear ODE (\ref{2gen}) to be of the form
\begin{eqnarray}            
\Omega=\lambda \frac{\partial}{\partial t}
+\mu \frac{\partial}{\partial x},
\label {sym03}
\end{eqnarray}
and
\begin{eqnarray}            
\Omega^{1}=\lambda \frac{\partial}{\partial t}
+\mu \frac{\partial}{\partial x}+(\dot{\mu}
-\dot{x}\dot{\lambda})\frac{\partial}{\partial \dot{x}},
\label {sym04}
\end{eqnarray}
respectively, where $\lambda$ and $\mu$ are the infinitesimals associated with the variables $t$ and $x$, respectively.  

\newtheorem{thm}{Theorem}
\begin{thm} 
Given the set of Lie point symmetries $\xi$ and $\eta$ of the linear ODE (\ref{horm1}), a set of nonlocal symmetries $\lambda$ and $\mu$ of the nonlinear ODE (\ref{2gen}) follows therefrom. 
\end{thm}
\emph {Proof :}
From the identity (\ref{gme02}) we define
\begin{eqnarray}            
\frac{\dot{U}}{U}=\frac{n\dot{x}}{x}+f(x)=X.\label{x}
\label {sym05}
\end{eqnarray}
The above relation is a contact type transformation  using which one can rewrite Eq. (\ref{2gen}) and Eq. (\ref{horm1}) as the Riccati equation
\bea
\dot{X}+X^2+c_1X+c_2=0.\qquad\quad\left(\dot{}=\frac{d}{dt}\right)
\label{reduced-riccati}
\eea
The symmetry vector field of this equation can be obtained by using the relation $X=\frac{\dot{U}}{U}$ and  rewriting Eq. (\ref{sym02}) as
\bea
&& \Lambda^{1}=\xi \frac{\partial}{\partial t}
+\bigg[\frac{\dot{\eta}}{U}-\frac{\eta \dot{U}}{U^2}
-X\dot{\xi}\bigg] \frac{\partial}{\partial X}\equiv\Sigma.
\label {sym06}
\eea
We note that Eq. (\ref{reduced-riccati}), being a first order ODE, admits infinite number of Lie point symmetries.  These Lie point symmetries of Eq. (\ref{reduced-riccati}) become contact symmetries of the linear second order ODE (\ref{horm1}) through the relation $X=\frac{\dot{U}}{U}$.

Similarly one can rewrite Eq. (\ref{sym04}) using the relation $X=n\frac{\dot{x}}{x}+f(x)$ as
\begin{eqnarray} 
&&\Omega^{1}=\lambda \frac{\partial}{\partial t}
+\bigg[(-\frac{n}{x^2}\dot{x}+f_x)\mu+(\dot{\mu}
-\dot{x}\dot{\lambda})\frac{n}{x}\bigg]\frac{\partial}{\partial X}\equiv\Xi,\qquad f_x=\frac{\partial f}{\partial x}.
\label {sym07}
\end{eqnarray}
As the symmetry vector fields $\Sigma$ and $\Xi$ are for the same
equation (\ref{reduced-riccati}), their infinitesimal symmetries must be equal.  Therefore, comparing equations (\ref{sym06}) and (\ref{sym07}) one obtains
\begin{eqnarray} 
&& \xi =\lambda,\quad 
 \bigg[\frac{\dot{\eta}}{U}-\frac{\eta \dot{U}}{U^2}
-f(x)\dot{\xi}\bigg]=\bigg[(-\frac{n}{x^2}\dot{x}
+f_x)\mu+\dot{\mu}\frac{n}{x}\bigg].
\label {sym08}
\end{eqnarray}
Rewriting the second equation in (\ref{sym08}) we arrive at the relation
\begin{eqnarray} 
 \frac{n}{x}\dot{\mu}+(-\frac{n}{x^2}\dot{x}+f_x)\mu= 
 \bigg[\frac{d}{dt}(\frac{\eta}{U})
-f(x)\dot{\xi}\bigg].
\label {sym09}
\end{eqnarray}

Since the infinitesimal symmetries $\xi$ and $\eta$ of the linear ODE are known and $U$ in (\ref{sym09}) is taken in the form (\ref{trans}), the right hand side now becomes an explicit function of $t$ and $x$. Solving the resultant first order linear ODE one can obtain the function $\mu$ which is nothing but the symmetry associated with the nonlinear ODE, see Appendix.  Since $U$ is given by the nonlocal form (\ref{trans}) the resultant symmetries in general turn out to be a nonlocal ones.  \hspace{12cm} $\square$

\subsection{Examples}
\noindent{\bf (a) Example 1:}

In order to illustrate the above theory we consider the simple parametric choice $c_1=c_2=0$ for which Eq. (\ref{2gen}) and Eq. (\ref{horm1}) reduce to the forms
\begin{eqnarray}
\ddot{x}+(n-1)\frac{\dot{x}^2}{x}+2\dot{x}f+\frac{1}{n}x\dot{x}f_x+\frac{x}{n}f^2=0,\qquad f_x=\frac{\partial f}{\partial x},\label{2gen2}
\end{eqnarray}
and
\begin{eqnarray}
\ddot{U}=0,\label{free}
\end{eqnarray}
respectively.  It is well known that the free particle equation (\ref{free}) admits the following eight Lie point symmetries, see for example \cite{hydon,ibragimov:1999,head:1993},
\begin{eqnarray} 
\Lambda_1=\frac{\partial}{\partial t},\quad            
\Lambda_2=\frac{\partial}{\partial U},\quad 
\Lambda_3=t \frac{\partial}{\partial U},\quad 
\Lambda_4=U\frac{\partial}{\partial U},\quad
\Lambda_5=U \frac{\partial}{\partial t},\nonumber\\ 
\Lambda_6=t \frac{\partial}{\partial t}
,\quad 
\Lambda_7=t^2 \frac{\partial}{\partial t}
+tU \frac{\partial}{\partial U},\quad 
\Lambda_8=tU \frac{\partial}{\partial t}
+U^2 \frac{\partial}{\partial U}.
\label {sym11}
\end{eqnarray}
Substituting the above symmetry generators $\Lambda_i$'s, $i=1,2,\ldots,8$, and $U=xe^{\int f(x)dt}$, in Eq. (\ref{sym09}),
we get $\xi=\lambda$ and the following seven first order ODEs for $\mu$,
\bea
&&\frac{n}{x}\dot{\mu}+(f_x-\frac{n}{x^2}\dot{x})\mu+(xf+n\dot{x})x^{-(n+1)}e^{-\int f(x)dt}=0,\\
&&\frac{n}{x}\dot{\mu}+(f_x-\frac{n}{x^2}\dot{x})\mu-(x-txf-nt\dot{x})x^{-(n+1)}e^{-\int f(x)dt}=0,\\
&&\frac{n}{x}\dot{\mu}+(f_x-\frac{n}{x^2}\dot{x})\mu=0,\\
&&\frac{n}{x}\dot{\mu}+(f_x-\frac{n}{x^2}\dot{x})\mu+(xf+n\dot{x})fx^{n-1}e^{-\int f(x)dt}=0,\\
&&\frac{n}{x}\dot{\mu}+(f_x-\frac{n}{x^2}\dot{x})\mu+f(x)=0,\\
&&\frac{n}{x}\dot{\mu}+(f_x-\frac{n}{x^2}\dot{x})\mu+2tf-1=0,\\
&&\frac{n}{x}\dot{\mu}+(f_x-\frac{n}{x^2}\dot{x})\mu+2(xf+n\dot{x})x^{2n-1}fe^{2\int f(x)dt}-1=0.
\eea
Integrating each one of the above first order linear ODEs we get the corresponding infinitesimal symmetry $\mu$.  Substituting the infinitesimal symmetries $\lambda$ and $\mu$ in (\ref{sym03}) we get following nonlocal symmetries of equation (\ref{2gen2}),
\begin{eqnarray} 
&&\hspace{-2cm}\Omega_1=\frac{\partial}{\partial t},\\
&&\hspace{-2cm}\Omega_2=\left(\frac{x^{-n}}{n}e^{\int (\frac{x}{n}f_x- f)dt}
-\frac{1}{n^2}
\int x^{1-n}f_xe^{\int(\frac{x}{n}f_x-f)dt}dt\right)
xe^{-\frac{1}{n}\int (x f_x)dt}
\frac{\partial}{\partial x},\label{eg-appendix}\\   
&&\hspace{-2cm}\Omega_3=\left(\frac{x^{-n}}{n}te^{\int (\frac{x}{n}f_x-f)dt}
-\frac{1}{n^2}\int tx^{1-n}f_xe^{\int(\frac{x}{n}f_x-f)dt}dt\right)xe^{-\frac{1}{n}\int (x f_x)dt}
\frac{\partial}{\partial x},\\    
&&\hspace{-2cm}\Omega_4=xe^{-\int(\frac{x}{n}f_x)dt}\frac{\partial}{\partial x},
\label{omega4-eg1}\\
%\eea
%\bea  
&&\hspace{-2cm}\Omega_5=x^{n}e^{\int f dt}\frac{\partial}{\partial t}
-\left(\frac{1}{n}\int x^{n-1}f(n\dot{x}+xf)e^{\int(\frac{x}{n}f_x+f) dt}dt\right)xe^{-\frac{1}{n}\int( x f_x) dt}
\frac{\partial}{\partial x},\\
%\eea
%\bea
&&\hspace{-2cm}\Omega_6=t\frac{\partial}{\partial t}
-xe^{-\frac{1}{n}\int( x f_x )dt}\left(\frac{1}{n}\int fe^{\frac{1}{n}\int (x f_x) dt}dt\right)  \frac{\partial}{\partial x},\\
&&\hspace{-2cm}\Omega_7=t^2 \frac{\partial}{\partial t}
+xe^{-\int{\frac{x}{n}f_x}dt}\left(\frac{1}{n}\int (1-2tf)e^{-\frac{1}{n}\int x f_xdt}dt\right)  \frac{\partial}{\partial x},\\
&&\hspace{-2cm}\Omega_8=tx^ne^{\int f dt}\frac{\partial}{\partial t}
+xe^{-\int{\frac{x}{n}f_x}dt}\left(\frac{1}{n}\int x^{n-1}(txf^2+n(tf-1)\dot{x})
e^{\int (\frac{1}{n}xf_x+f)dt}dt \right)
\frac{\partial}{\partial x}.
\label {sym12}
\end{eqnarray}
One can verify that each one of the above nonlocal symmetries indeed satisfies the invariance condition (\ref{invariance}) and is the nonlocal symmetry vector field of (\ref{2gen2}).  This is demonstrated in the Appendix for a particular symmetry vector, namely $\Omega_2$, as an example.  

We note here that there is also a possibility to find the nonlocal symmetries of Eq. (\ref{2gen2}) by  introducing suitable auxiliary/covering equation and deducing the point symmetries associated with the combined system giving rise to a one parameter group as studied in Refs. \cite{krasilshchik1,krasilshchik2}.  However, we have not explored such a possibility here.

\n{\bf Proposition 1:} \emph{The nonlocal symmetry $\Omega_4$ reduces Eq. (\ref{2gen2}) to the Riccati equation
$dz/dt=-nz^2$ through the reduction transformation $z=\frac{\dot{x}}{x}+\frac{f}{n}$.}
%\bea
%z=\frac{\dot{x}}{x}+\frac{f}{n}.\label{characteristic1}
%\eea

%\n Integrating the reduced Riccati equation we get the following general solution
%\bea
%z=\frac{1}{I_1+nt},
%\eea
%where $I_1$ is the integration constant.  
\noindent\emph{Proof :} Let us consider  the Lagrange's system associated with the nonlocal symmetry $\Omega_4$ given by Eq. (\ref{omega4-eg1}),
\bea
\frac{dt}{0}=\frac{dx}{x}=\frac{d\dot{x}}{\dot{x}-\frac{x^2}{n}f_x}.  
\eea
The characteristics are $t$ and 
\bea
z=\frac{\dot{x}}{x}+\frac{f}{n}.\label{characteristic1}
\eea
We find the reduced equation of (\ref{2gen2}) is the following Riccati equation,
\bea
\frac{dz}{dt}=-nz^2,
\eea
whose general solution is given as
\bea
z=\frac{1}{I_1+nt},
\eea
where $I_1$ is the integration constant.  \hspace{8cm}$\square$

Substituting the above solution in (\ref{characteristic1}) and rearranging we get
\bea
\dot{x}-\frac{x}{I_1+nt}+x\frac{f}{n}=0\label{riccati-eg1}
\eea
Solving the above equation one can find the general solution of (\ref{2gen2}).  However, one finds that Eq. (\ref{riccati-eg1}) can be integrated only for certain specific forms of $f$.  One such form of $f$ for which Eq. (\ref{riccati-eg1}) is integrable is $f=kx^m$.  For this choice of $f$ and $n=1$, Eq. (\ref{2gen2}) reduces to the generalized Emden equation \cite{feix:97:01,gladwin:jmp:2009,chandrasekar:06:01:jpa}
\begin{eqnarray}
\ddot{x}+(m+2)kx^m\dot{x}+k^2x^{2m+1}=0,
\label{eg1}
\end{eqnarray}
whose general solution is obtained by integrating  (\ref{riccati-eg1}) as
\bea
x(t)=\frac{I_1+t}{\left[I_2+\frac{km}{m+1}(I_1+t)^{m+1}\right]^{\frac{1}{m}}}.\label{eg1sol}
\eea
where $I_1$ and $I_2$ are the integration constants, which agrees with the known result \cite{chandrasekar:06:01:jpa}.  We wish to point out here that in additon to the above nonlocal symmetries,  Eq. (\ref{eg1}) has the following Lie point symmetries which can be deduced using the standard procedure, for example using MULIE \cite{head:1993,hydon,ibragimov:1999},
\bea
\Omega_1=\frac{\partial}{\partial t},\qquad \Omega_{9}=t\frac{\partial}{\partial t}-\frac{x}{m}\frac{\partial}{\partial x}.
\eea
Obviously the symmetries $\Omega_9$ is outside the scope of the above nonlocal connection (Theorem 1).

\noindent{\bf (b) Example 2:}

Next we consider another interesting nonlinear ODE of the form
\begin{eqnarray}
\ddot{x}+c_2x+\frac{k^2}{x^3}=0,\label{example2}
\end{eqnarray}
which arises in different areas of physics and has been studied in  Ref. \cite{pinney,chandrasekar:06:01:jpa,euler:2007,ermakov,hawkins}.  This equation arises in a wide variety of fields such as the study of cosmological field \cite{ermakov-app1}, quantum field theory in curved space \cite{ermakov-app2}, quantum cosmology \cite{ermakov-app3}, molecular structures \cite{ermakov-app4,ermakov-app5} and Bose-Einstein condensation \cite{rajendran}.
Equation (\ref{example2}) is found to be connected to the harmonic oscillator equation
\begin{eqnarray}
\ddot{U}+c_2U=0,\label{sho}
\end{eqnarray}
by the nonlocal transformation $U=xe^{\int \frac{k}{x^2} dt}$.  The nonlocal symmetries associated with Eq. (\ref{example2}) can be found by following the procedure discussed in Sec. \ref{procedure}.  Substituting the following known Lie point symmetries of the harmonic oscillator (\ref{sho}) in (\ref{sym09}) \cite{hydon,ibragimov:1999,head:1993},
\begin{eqnarray}
&&\hspace{-2cm}\Lambda_1=\frac{\partial}{\partial t},\,\,\Lambda_2=\sin 2\omega t\frac{\partial}{\partial t}+\omega U\cos2\omega t\frac{\partial}{\partial U},\,\,\Lambda_3=\cos2\omega t\frac{\partial}{\partial t}-\omega U\sin2\omega t\frac{\partial}{\partial U},\nonumber\\
&&\hspace{-2cm}\Lambda_4=U(\sin\omega t\frac{\partial}{\partial t}+\omega U\cos\omega t\frac{\partial}{\partial U}),\,\,\Lambda_5=U\frac{\partial}{\partial U},\,\,\Lambda_6=U(\cos\omega t\frac{\partial}{\partial t}-\omega U\sin\omega t\frac{\partial}{\partial U}),\nonumber\\
&&\hspace{-2cm}\Lambda_7=\sin\omega t\frac{\partial}{\partial U},\,\,\Lambda_8=\cos\omega t\frac{\partial}{\partial U},
\end{eqnarray}
where $\omega=\sqrt{c_2}$,
we get a set of first order ODEs.  Solving these first order ODEs, with the substitution $U=xe^{\int\frac{k}{x^2}dt}$, we get the following nonlocal symmetries of Eq. (\ref{example2}):
\begin{eqnarray}
&&\hspace{-1.5cm} \Omega_1=\frac{\partial}{\partial t},\\
&&\hspace{-1.5cm}\Omega_2=\sin2\omega t\frac{\partial}{\partial t}-2\omega xe^{2k\int \frac{1}{x^2}dt}\int\alpha_1e^{-2k\int \frac{1}{x^2}dt} dt)\frac{\partial}{\partial x},\\
&&\hspace{-1.5cm}\Omega_3=\cos2\omega t\frac{\partial}{\partial t}+2\omega xe^{2k\int \frac{1}{x}dt}\left(\int \alpha_2e^{-2k\int \frac{1}{x^2}} dt\right)\frac{\partial}{\partial x},\\
&&\hspace{-1.5cm}\Omega_4=xe^{k\int \frac{1}{x^2}dt}\left[\sin\omega t\frac{\partial}{\partial t}+e^{k\int\frac{1}{x^2}dt}\int \alpha_3\left(\frac{e^{-k\int \frac{1}{x^2}dt}}{x}\right)dt\frac{\partial}{\partial t}\right],\\
&&\hspace{-1.5cm}\Omega_5=xe^{2k\int \frac{1}{x^2}dt}\frac{\partial}{\partial x},\label{omega5-eg2}\\
&&\hspace{-1.5cm}\Omega_6=xe^{k\int\frac{1}{x^2}dt}\left[\cos\omega t\frac{\partial}{\partial t}-e^{k\int\frac{1}{x^2}dt}\int\bigg(k^2\cos(\omega t)+x^4\omega^2\cos(\omega t)\right.\nn\\
&&\qquad\qquad\left.+x\dot{x}\cos(\omega t)+w x^3\dot{x}\sin(\omega t)\bigg)
\frac{e^{-k\int\frac{1}{x^2}dt}}{x^3}dt\frac{\partial}{\partial x}\right],\\
&&\hspace{-1.5cm}\Omega_7=xe^{2k\int \frac{1}{x^2}dt}\left[
\int (\frac{\omega}{x}\cos(\omega t)-\frac{\dot{x}}{x^2}\sin(\omega t)%\right.\nonumber\\
%&&\qquad\qquad\left.
-\frac{k}{x^3}\sin(\omega t))e^{-3k\int \frac{1}{x^2}dt}dt\right]\frac{\partial}{\partial x},\\
&&\hspace{-1.5cm}\Omega_8=xe^{2k\int \frac{1}{x^2}dt}\left[\int \frac{1}{x}e^{-\int\frac{3k}{x^2}dt}\left(\omega\sin(\omega t)+(\frac{k}{x^2}+\frac{\dot{x}}{x})\cos(\omega t)\right)  dt\right]\frac{\partial}{\partial x},\,\,\,\,
\end{eqnarray}
where
\begin{eqnarray}
&&\hspace{-2cm}\alpha_1=\omega\sin2\omega t+\frac{k}{x^2}\cos2\omega t,\nonumber\\
&&\hspace{-2cm}\alpha_2=\frac{k}{x^2}\sin2\omega t-\omega\cos2\omega t,\nonumber\\
&&\hspace{-2cm}\alpha_3=\omega x\dot{x}\cos(\omega t)-\frac{x\dot{x}}{x}\sin(\omega t)-\omega^2x^2\sin(\omega t)-\frac{k^2}{x^2}\sin(\omega t),\nn
\end{eqnarray}
We also wish to point out here that in addition to the above nonlocal symmetries,  Eq. (\ref{example2}) has the following three Lie point symmetries,
\bea
&&\Omega_9=\left(2\sin^2(\sqrt{c_2}t)-1\right)\frac{\partial}{\partial t}+2\sqrt{c_2}x\cos(\sqrt{c_2}t)\sin(\sqrt{c_2}t)\frac{\partial}{\partial x},\\
&&\Omega_{10}=2\cos(\sqrt{c_2}t)\sin(\sqrt{c_2}t)\frac{\partial}{\partial t}+\left(2\sqrt{c_2}x\sin^2(\sqrt{c_2}t)-\sqrt{c_2}x\right)\frac{\partial}{\partial x},\\
&&\Omega_{11}=\frac{1}{c_2}\left(\sin^2(\sqrt{c_2}t)-1\right)\frac{\partial}{\partial t}-\frac{\sqrt{x}}{\sqrt{c_2}}\sin(\sqrt{c_2}t)\cos(\sqrt{c_2}t)\frac{\partial}{\partial x}.
\eea

\n{\bf Proposition 2:} \emph{The nonlocal symmetry $\Omega_5$ reduces Eq. (\ref{example2}) to the Riccati equation
$dz/dt=-z^2+c_2$ through the reduction transformation $z=\dot{x}/x+k/x^2$.}

\noindent\emph{Proof :} Let us consider the Lagrange's system associated with the nonlocal symmetry $\Omega_5$ given by Eq. (\ref{omega5-eg2}),
\bea
\frac{dt}{0}=\frac{dx}{x}=\frac{d\dot{x}}{\dot{x}+\frac{2k}{x^3}}
\eea
The characteristics of this system are $t$ and 
\bea
z=\frac{\dot{x}}{x}+\frac{k}{x^2}.  \label{character-z1}
\eea
The reduced equation of (\ref{example2}) is found to be 
\bea
\frac{dz}{dt}=-z^2+c_2.
\eea
The general solution of the above Riccati equation is
\bea
z=-\sqrt{c_2}\tan(\sqrt{c_2}(t-I_1)).
\eea
\hspace{15cm}$\square$

\noindent Substituting the expression for $z$ in (\ref{character-z1})  and rearranging we get
\bea
\dot{x}+\sqrt{c_2}x\tan(\sqrt{c_2}t+I_1))+\frac{k}{x}=0.\label{eg2-sol1}
\eea
%\bea
%z=\frac{\dot{x}}{x}+\frac{k}{x^2}.  \label{character-z1}
%\eea
%The reduced equation of (\ref{example2}) is found to be 
%\bea
%\frac{dz}{dt}=-z^2+c_2.
%\eea
%The general solution of the above Riccati equation is
%\bea
%z=-\sqrt{c_2}\tan(\sqrt{c_2}(t-I_1)).
%\eea
Integrating (\ref{eg2-sol1}) we find the general solution of (\ref{example2}) as
\bea
x(t)=\frac{\cos(\sqrt{c_2}t+I_1)}{\left(I_2-\frac{2k}{\sqrt{c_2}}\tan(\sqrt{c_2}t+I_1)\right)^{\frac{1}{2}}},
\eea
where $I_1$ and $I_2$ are the integration constants, which agrees with the known solution \cite{chandrasekar:06:01:jpa}.  
\subsection{Extension to more general class of second order ODEs}
\label{sub-sec-secondorder}
The procedure described to deduce the nonlocal symmetries of Eq. (\ref{2gen}) can be further extended to deduce the nonlocal symmetries of more general nonlinear ODEs of the form
\bea
\left(D_h^2+c_1(t)D_h+c_2(t)\right)g(x,t)=0,\label{mgen1}
\eea
where $D_h=\left(\frac{d}{dt}+f(x,t)\right)$.  The above equation (\ref{mgen1}) is related to the following nonautonomous linear ODE
\bea
\ddot{U}+c_1(t)\dot{U}+c_2(t)U=0,\label{timedeplinear} 
\eea
through the nonlocal transformation
\bea
U=g(x,t)e^{\int f(x,t)dt}.\label{arbitrary-nonlocal-trans}
\eea
\begin{thm} 
Equation (\ref{mgen1}) admits a set of nonlocal symmetries which can be obtained directly from the Lie point symmetries $\xi$ and $\eta$ of Eq. (\ref{timedeplinear}).
\end{thm}

\noindent\emph{ Proof :}
From the above nonlocal transformation we define 
\bea
X=\frac{\dot{U}}{U}=\frac{\dot{g}}{g}+f(x,t).\label{arbitrary-contact}
\eea
The above relation is a contact type transformation  using which one can rewrite Eq. (\ref{mgen1}) and Eq. (\ref{timedeplinear}) as the Riccati equation
\bea
\dot{X}+X^2+c_1(t)X+c_2(t)=0,\qquad\quad\dot{}=\frac{d}{dt}.
%\label{reduced-riccati}
\eea
The symmetry vector field of this equation can be again obtained by using the relation $X=\frac{\dot{U}}{U}$ and  rewriting Eq. (\ref{sym02}) and Eq. (\ref{sym04}) we get
\bea
&&\hspace{2cm}\Lambda^1=\xi\frac{\partial}{\partial t}+\left[\frac{\dot{\eta}}{U}-\frac{\eta\dot{U}}{U^2}-X\dot{\xi}\right]\frac{\partial}{\partial X}\equiv\Sigma\qquad\qquad \nonumber %(\ref{sym06})
\eea
and
\bea
\hspace{-1.2cm}\Omega^1=\lambda\frac{\partial}{\partial t}+\left\{(\dot{\mu}-\dot{x}\dot{\lambda})\frac{g_x}{g}+\mu\left[f_x+\frac{\partial}{\partial x}\left(\frac{\dot{g}}{g}\right)\right]+\lambda\frac{d}{dt}\left(\frac{ g_t}{g}\right)+\lambda f_t\right\}\frac{\partial}{\partial X}\equiv\Xi,\label{sym-gen}
\eea
respectively.  Comparing (\ref{sym06}) and (\ref{sym-gen}) we get
\bea
\hspace{-1.2cm}\xi=\lambda,\quad \frac{\dot{\eta}}{U}-\frac{\eta\dot{U}}{U^2}-f(x,t)\dot{\xi}=\dot{\mu}\frac{g_x}{g}+\mu\left\{f_x+\frac{\partial}{\partial x}\left(\frac{\dot{g}}{g}\right)\right\}+\frac{d}{dt}\left(\frac{\xi g_t}{g}\right)+\xi f_t.\label{sym-gen2}
\eea
Rewriting the second equation in (\ref{sym-gen2}) we arrive at the relation
\bea
\frac{g_x}{g}\dot{\mu}+\left\{f_x+\frac{\partial}{\partial x}\left(\frac{\dot{g}}{g}\right)\right\}\mu=\frac{d}{dt}\left(\frac{\eta}{U}\right)-f(x,t)\dot{\xi}-\frac{d}{dt}\left(\frac{\xi g_t}{g}\right)-\xi f_t.\label{arbitrary-sym}
\eea
Substituting $U=g(x,t)e^{\int f(x,t)dt}$ and the point symmetries of Eq. (\ref{timedeplinear}) in the above first order linear ODE and solving for $\mu$, we can obtain a set of nonlocal symmetries of Eq. (\ref{mgen1}).\hspace{14cm}$\square$

\section{Nonlocal symmetries : Third order ODEs}
\label{third-order-symmetry}
The procedure to deduce nonlocal symmetries discussed in Sec. \ref{procedure} can be straightforwardly extended to third order ODEs. 
In this section we use this procedure to deduce the nonlocal symmetries of the following class of third order nonlinear ODEs,
\bea
&&\hspace{-1cm}\dddot{x}+\frac{1}{n}((3nf+xf_x+3n(n-1)\frac{\dot{x}}{x})\ddot{x})+(n-1)(n-2)\frac{\dot{x}^3}{x^2}\nonumber\\
&&\hspace{-1cm}\qquad+(x(3nf_x+xf_{xx})+3n(n-1)f)\frac{\dot{x}^2}{nx}+3(nf+xf_x)f\frac{\dot{x}}{n}
+\frac{x}{n}f^3=0.\label{thirdordernonlineareq}
\eea
\begin{thm}
A class of nonlocal symmetries of the nonlinear ODE (\ref{thirdordernonlineareq}) can be obtained directly from the Lie point symmetries of the second order linear ODE
\bea
\dddot{U}=0,\label{thirdorderlineareq}
\eea
\end{thm}
\emph{Proof :}
It is straightforward to check that Eqs. (\ref{thirdordernonlineareq}) are (\ref{thirdorderlineareq}) and connected through the nonlocal transformation 
\bea
U=x^ne^{\int f(x)dt}.
\eea
From the above relation we find $\frac{\dot{U}}{U}=\frac{\dot{x}}{x}+f$, which is same as Eq. (\ref{x}).  Therefore, we find that the procedure discussed in Sec. \ref{procedure} can be straightforwardly applied to Eq. (\ref{thirdordernonlineareq}) as well and the nonlocal symmetries are obtained by substituting the Lie point symmetries of (\ref{thirdorderlineareq}) in (\ref{sym09}) and solving the resultant equations.\hspace{6cm}$\square$

The third order linear ODE (\ref{thirdorderlineareq}) is known to admit the following seven Lie point symmetries \cite{hydon,head:1993,ibragimov:1999},
\begin{eqnarray}
&& \Lambda_1=\frac{\partial}{\partial t},\,\,\,\Lambda_2=\frac{\partial}{\partial U},\,\,\,\Lambda_3=t^2\frac{\partial}{\partial U},\,\,\,\Lambda_4=t\frac{\partial}{\partial t},\,\,\,\Lambda_5=t\frac{\partial}{\partial U},\,\,\,\Lambda_6=U\frac{\partial }{\partial U},\nonumber\\
&&\Lambda_7=\frac{t^2}{2}\frac{\partial}{\partial t}+Ut\frac{\partial}{\partial U}.
\end{eqnarray}
Substituting the above Lie point symmetry vector fields in Eq. (\ref{sym09}) and solving the resultant first order linear ODEs we find the following symmetry vector fields of Eq. (\ref{thirdordernonlineareq}),
\bea
&&\Omega_1=\frac{\partial}{\partial t},\quad\Omega_2=-xe^{-\int \frac{xf_x}{n}dt}\int\left( x^{-(n+1)}(\frac{x}{n}f+\dot{x})e^{\int(\frac{xf_x}{n}-f)dt}\right)dt
\frac{\partial}{\partial x},\\
&&\Omega_3=xe^{-\int\frac{xf_x}{n}dt}\int \frac{1}{nx^2}e^{\int(\frac{xf_x}{n}-f) dt}\left(2t-t^2f-2t^2\frac{\dot{x}}{x}\right)dt\frac{\partial}{\partial x},\\
&&\Omega_4=t\frac{\partial}{\partial t}-xe^{-\int \frac{xf_x}{n}dt}\int\left(\frac{f}{n}e^{\int(\frac{xf_x}{n})dt}\right)dt\frac{\partial}{\partial x},\\
&&\Omega_5=xe^{-\int \frac{xf_x}{n}}\int e^{\int\frac{1}{nx^2}( \frac{xf_x}{n}-f)dt}\left(1-tf-2t\frac{\dot{x}}{x}\right)dt\frac{\partial}{\partial x},\\
&&\Omega_6=xe^{-\int \frac{xf_x}{n}dt}\frac{\partial}{\partial x},\label{omega6}\\
&&\Omega_7=\frac{t^2}{2}\frac{\partial}{\partial t}+\frac{x}{n}e^{-\int \frac{xf_x}{n}dt}\int (1-tf)e^{\int(\frac{xf_x}{n})dt}dt\frac{\partial}{\partial x}.
\eea
\n{\bf Proposition 3:} \emph{The nonlocal symmetry $\Omega_6$ reduces Eq. (\ref{thirdordernonlineareq}) to the modified Emden equation/second order Riccati equation
$\frac{d^2z}{dt^2}+3nz\dot{z}+n^2z^3$ through the reduction transformation $z=\dot{x}/x+f/n$.}

To check the above assertion let us consider the Lagrange's system associated with the symmetry $\Omega_6$ given by Eq. (\ref{omega6}),
\bea
\frac{dt}{0}=\frac{dx}{x}=\frac{d\dot{x}}{\dot{x}-x^2\frac{f_x}{n}}.
\eea
The characteristics of this system are $t$ and 
\bea
z=\frac{\dot{x}}{x}+\frac{f}{n}.\label{character-z2}
\eea
  The reduced equation now turns out to be of the form
\bea
\frac{d^2z}{dt^2}+3nz\dot{z}+n^2z^3=0,\label{mee}
\eea
which is the modified Emden equation and also known as the second order Riccati equation.  Note here that eq. (\ref{eg1}) reduces to  Eq. (\ref{mee}) for the choice $m=1$.  The solution of (\ref{mee}) can be therefore obtained from (\ref{eg1sol}) with the substitution $m=1$ and is given as
\bea
z=\frac{I_1+t}{(I_2+I_1t+\frac{t^2}{2})},
\eea
where $I_1$ and $I_2$ are integration constants.\hspace{8cm}$\square$

Rearranging the reduction transformation with the substitution $z=(I_1+t)/((I_2+I_1t+\frac{t^2}{2}))$, we get
\bea
\dot{x}-\frac{x(I_1+t)}{(I_2+I_1t+\frac{t^2}{2})}+\frac{x}{n}f=0.\label{thirdordersoleq}
\eea
We note that the above equation is integrable only for certain specific forms of $f$.  We consider one such simple form for $f$ as $f=kx$.   For this form of $f$ Eq. (\ref{thirdordernonlineareq}) reduces to a special case of the Chazy equation XII \cite{chazy:11:01,euler:2004,cosgrove:00:01,halburd:1999,chandrasekar:06:01:jpa},
\begin{eqnarray}
\dddot{x}+4kx\ddot{x}+3k\dot{x}^2+6k^2x^2\dot{x}+k^3x^4=0.\label{egthirdorder}
\end{eqnarray}
Integrating Eq. (\ref{thirdordersoleq}) with $f=kx$, we get the general solution of (\ref{egthirdorder}) as
\bea
x(t)=\frac{\frac{kt^2}{2}+I_1t+I_1I_2}{I_1I_3+kI_1I_2t+\frac{kI_1}{2}t^2+\frac{k^2t^3}{6}},
\eea
where $I_1,\,I_2$ and $I_3$ are the integration constants. We wish to note that, in addition to the above nonlocal symmetries, Eq. (\ref{egthirdorder}) possesses the following Lie point symmetries also,
\bea
&&\Omega_1=\frac{\partial}{\partial t},\quad \Omega_8= x\frac{\partial}{\partial x}-t\frac{\partial}{\partial t},\quad \Omega_{9}=
-\frac{t^2}{2}\frac{\partial}{\partial t}+xt\frac{\partial }{\partial x}-\frac{3}{2k}\frac{\partial}{\partial x}.
\eea
\subsection{More general class of third order ODEs}
In addition to Eq. (\ref{thirdordernonlineareq}) one finds a more general class of third order ODEs of the following form
\bea
(D_h^3+c_1(t)D_h^2+c_2(t)D_h+c_3(t))g(x,t)=0,\label{gen-thirdorder}
\eea
where $D_h=\left(\frac{d}{dt}+f(x,t)\right)$ and $c_i(t),\,i=1,2,3$, are arbitrary functions of $t$, which admits nonlocal symmetries.  
This class of third order nonlinear ODEs is related to the following nonautonomous third order linear ODE of the form
\bea
\dddot{U}+c_1(t)\ddot{U}+c_2(t)\dot{U}+c_3(t)U=0,\label{linear-third}
\eea
through the nonlocal transformation $U=g(x,t)e^{\int f(x,t)dt}$.  In order to identify the nonlocal symmetries associated with Eq. (\ref{gen-thirdorder}), one can straightforwardly apply the procedure discussed in Sec. \ref{sub-sec-secondorder}.  Substituting the point symmetries $\xi$ and $\eta$ of the third order linear ODE (\ref{sym-gen2}) and solving, one can obtain the nonlocal symmetries of (\ref{gen-thirdorder}).

\section{Arbitrary order nonlinear ODEs}
\label{arbitrary-order}
Having discussed the applicability of the procedure to obtain the nonlocal symmetries of certain class of second and third order ODEs we extend the procedure to a class of arbitrary order nonlinear ODEs.  In this context the following theorem holds good.
\begin{thm}
A set of nonlocal symmetries of the $m^{th}$ nonlinear ODE 
\bea
\left(D_h^m+c_1(t)D_h^{m-1}+\ldots+c_{m-1}(t)\right)g(x,t)=0,\label{gen-arbitrary}
\eea
where $D_h^m=\left(\frac{d}{dt}+f(x,t)\right)^m$, 
can be obtained directly from the Lie point symmetries of the $m^{th}$ order linear ODE
\bea
U^{(m)}+c_1(t)U^{(m-1)}+\ldots+c_{m-1}(t)U=0,\qquad U^{(m)}=\frac{d^mU}{dt^m}.\label{gen-arbitrarylinear}
\eea
\end{thm}
\emph{Proof :}
The nonlinear ODE (\ref{gen-arbitrary}) is connected to the linear ODE (\ref{gen-arbitrarylinear}) through the nonlocal transformation $U=g(x,t)e^{\int f(x,t)dt}$.  Note that this nonlocal transformation is the same as (\ref{arbitrary-nonlocal-trans}), connecting the second order linear ODE (\ref{timedeplinear}) and the nonlinear ODE (\ref{mgen1}).  Consequently, a set of nonlocal symmetries of Eq. (\ref{gen-arbitrary}) can be found in principle by substituting the point symmetries of the linear ODE (\ref{gen-arbitrarylinear}) in Eq. (\ref{sym-gen2}) and solving the resultant equations, as in the case of second and third order nonlinear ODEs. \hspace{1cm}$\square$

However, we note here that one cannot obtain all the point symmetries of the linear ODE  (\ref{gen-arbitrarylinear}) of arbitrary order $m$.  Therefore we consider a specific parametric choice 
 $c_i(t)=0,\,i=1,2,\ldots,m-1$, which reduces Eq. (\ref{gen-arbitrarylinear}) to the form
\bea
\frac{d^mU}{dt^m}=0.\label{arbitrarylinear}
\eea
Equation (\ref{arbitrarylinear}) admits at least the following two point symmetries for arbitrary order $m$,  
\bea
\Lambda_1=\frac{\partial}{\partial t},\quad\Lambda_2=U\frac{\partial }{\partial U}.
\eea
Substituting now the nonlocal transformation $U=g(x,t)e^{\int f(x,t)dt}$ in (\ref{arbitrarylinear}) we get the nonlinear ODE
\bea
\left(\frac{d}{dt}+f(x,t)\right)^mg(x,t)=0.\label{arbitrarynonlinear}
\eea
Note that Eq. (\ref{arbitrarynonlinear}) is a generalization of Eqs. (\ref{2gen2}) and (\ref{thirdordernonlineareq}).  To identify the nonlocal symmetries of (\ref{arbitrarynonlinear}), we substitute the Lie point symmetries of (\ref{arbitrarylinear}) in (\ref{sym-gen2}). Solving the resultant equations we deduce the following nonlocal symmetries of the nonlinear ODE (\ref{arbitrarynonlinear}), 
\bea
&&\hspace{-1.5cm}\Omega_1=\frac{\partial}{\partial t}-\left[e^{-\int \frac{g}{g_x}(f_x+\frac{\partial}{\partial x}(\frac{\dot{g}}{g}))dt}\int \frac{g}{g_x}\left[\frac{d}{dt}+f_t\right]\left(\frac{g_t}{g}\right)e^{\int \frac{g}{g_x}(f_x+\frac{\partial}{\partial x}\left(\frac{\dot{g}}{g}\right))dt}dt\right]\frac{\partial}{\partial x},\label{nonlocal-gen-symmetry1}\\
&&\hspace{-1.5cm}\Omega_2=\exp\left[-\displaystyle\int\left\{\frac{g}{g_x}\left[f_x+\frac{\partial}{\partial x}\left(\frac{\dot{g}}{g}\right)\right]\right\}dt\right]\frac{\partial}{\partial x}.\hspace{4cm}\label{nonlocal-gen-symmetry}
\eea

Let us consider the specific choice $g(x,t)=x^n$, $f(x,t)=f(x)$, which reduces Eq.(\ref{gen-arbitrary}) to the following form,
\bea
\left(\frac{d}{dt}+f(x)\right)^m x^n=0.\label{arbitrarynonlinear1}
\eea
The above equation is a generalization of the Riccati and Abel chains.  The nonlocal symmetries associated with this equation is obtained by substituting $g(x,t)=x^n$ and $f(x,t)=f(x)$ in (\ref{nonlocal-gen-symmetry1}) and  (\ref{nonlocal-gen-symmetry}) and are given as
\bea
\Omega_1=\frac{\partial}{\partial t},\qquad\quad\Omega_2=xe^{-\int \frac{x}{n}f_xdt}\frac{\partial}{\partial x}.
\eea

\n{\bf Proposition 4:} \emph{The nonlocal symmetry $\Omega_2$ reduces Eq. (\ref{arbitrarynonlinear1}) to the integrable Riccati chain
$\left(\frac{d}{dt}+nz\right)^{m-1}z=0$ through the reduction transformation $z=\frac{\dot{x}}{x}+\frac{f}{n}$.}

\n\emph{Proof :}
\n Consider now the Lagrange's system associated with $\Omega_2$ which is
\bea
\frac{dt}{0}=\frac{dx}{x}=\frac{d\dot{x}}{\dot{x}-\frac{x^2}{n}f_x}.
\eea
The characteristics are $t$ and 
\bea
z=\frac{\dot{x}}{x}+\frac{f}{n}.\label{character-z3}
\eea  
The reduced equation is then found to be
\bea
\left(\frac{d}{dt}+nz\right)^{m-1}z=0.\label{riccati}
\eea
\hspace{15cm}$\square$

We know that the Riccati chain can be integrated to get the general solution \cite{chandrasekar:06:01:jpa} for a specified order $m$, say, $z=v(t)$.
 Substituting this in the reduction transformation and rearranging we get
\bea
\dot{x}-v(t)x+\frac{x}{n}f(x)=0.\label{arbitrary-first-order}
\eea
One can obtain the general solution of (\ref{arbitrarynonlinear1}) by solving the above first order nonlinear ODE.  Thus we find that the problem of solving any arbitrary equation belonging to the class (\ref{arbitrarynonlinear1}) is reduced to solving the first order ODE (\ref{arbitrary-first-order}).
\section{Coupled Second order nonlinear ODEs}
\label{coupled-symmetry}
Having discussed the procedure for deducing the nonlocal symmetries for a class of arbitrary order ODE, we now extend the procedure to coupled second order ODEs.
Let us consider the following system of coupled second order ODEs,
%\begin{subequations}
\begin{numparts}
\begin{eqnarray}
\addtocounter{equation}{-1}
\label{gencoupledeq}
\addtocounter{equation}{1}
&&\ddot{x}+(n-1)\frac{\dot{x}^2}{x}+2\dot{x}f+\frac{x}{n}(f_x\dot{x}+f_y\dot{y})+\frac{x}{n}f^2=0,\\
&&\ddot{y}+(n-1)\frac{\dot{y}^2}{y}+2\dot{y}g+\frac{y}{n}(g_x\dot{x}+g_y\dot{y})+\frac{y}{n}g^2=0,
\end{eqnarray}
\end{numparts}
%\end{subequations}
$f_x=\frac{\partial f}{\partial x}$, $g_x=\frac{\partial g}{\partial x}$, $f_y=\frac{\partial f}{\partial y}$ and $g_y=\frac{\partial g}{\partial y}$, which are related to the system of free particle equations 
\bea
\ddot{U}=0,\qquad\qquad\ddot{V}=0,\label{2dlinear}
\eea
through the nonlocal transformations
\bea
U=x^ne^{\int f(x,y)dt},\qquad\qquad V=y^ne^{\int g(x,y)dt}.\label{nonlocalcoupledtrans}
\eea
Equation (\ref{gencoupledeq}) includes the coupled modified Emden equation \cite{gladwin:jpa:2009} and the coupled generalized Duffing-van der Pol oscillator equation for specific forms of $f$ and $g$.  The integrability of Eq. (\ref{gencoupledeq}) and its further generalizations have been studied in \cite{gladwin:jmp:2010}.
The symmetry vector field associated with the system of linear equation (\ref{2dlinear}) is given by
\bea
\Lambda=\xi\frac{\partial}{\partial t}+\eta_1\frac{\partial}{\partial U}+\eta_2\frac{\partial}{\partial V}.
\eea
The first prolongation of this vector field is
\bea
\Lambda^1=\xi\frac{\partial}{\partial t}+\eta_1\frac{\partial}{\partial U}+\eta_2\frac{\partial}{\partial V}+(\dot{\eta}_1-\dot{U}\dot{\xi})\frac{\partial}{\partial \dot{U}}+(\dot{\eta}_2-\dot{V}\dot{\xi})\frac{\partial}{\partial \dot{V}}.\label{linearprolongation}
\eea
We assume that the system of nonlinear equations (\ref{gencoupledeq}) admits a symmetry vector field of the form
\bea
\Omega=\lambda\frac{\partial}{\partial t}+\mu_1\frac{\partial}{\partial x}+\mu_2\frac{\partial}{\partial y}
\eea
and its prolongation is given as
\bea
\Omega^1=\lambda\frac{\partial}{\partial t}+\mu_1\frac{\partial}{\partial x}+(\dot{\mu}_1-\dot{x}\dot{\lambda})\frac{\partial}{\partial \dot{x}}+(\dot{\mu}_2-\dot{y}\dot{\lambda})\frac{\partial}{\partial \dot{y}}.\label{nonlinearprolongation}
\eea
\begin{thm} 
A set of nonlocal symmetries $\lambda$, $\mu_1$ and $\mu_2$ of Eq. (\ref{gencoupledeq}) for the case $f=g$ can be obtained from the point symmetries $\xi$, $\eta_1$ and $\eta_2$ of Eq. (\ref{2dlinear}).
\end{thm}
\emph{Proof :}
Using the nonlocal transformations (\ref{nonlocalcoupledtrans}) one can write the following identities,
\bea
\frac{\dot{U}}{U}=n\frac{\dot{x}}{x}+f(x,y)=X,\qquad \frac{\dot{V}}{V}=n\frac{\dot{y}}{y}+g(x,y)=Y.
\eea
Using the above contact transformations, one can rewrite Eqs. (\ref{gencoupledeq}) and (\ref{2dlinear}) in terms of the new variables $X$ and $Y$.  The symmetry vector field of these new equations can be obtained by using the relations $X=\frac{\dot{U}}{U}$, $Y=\frac{\dot{V}}{V}$ and rewriting (\ref{linearprolongation}) as
\bea
&&\hspace{-0.9cm}\Lambda^1=\xi\frac{\partial}{\partial t}+\left[\frac{\dot{\eta}_1}{U}-\eta_1\frac{\dot{U}}{U^2}-X\dot{\xi}\right]\frac{\partial}{\partial X}+\left[\frac{\dot{\eta}_2}{V}-\eta_2\frac{\dot{V}}{V^2}-Y\dot{\xi}\right]\frac{\partial}{\partial Y}\equiv\Sigma.
\eea
Similarly one can rewrite (\ref{nonlinearprolongation}) using the relation $X=\frac{n\dot{x}}{x}+f(x,y)$ and $Y=\frac{n\dot{y}}{y}+g(x,y)$ as
\bea
&&\hspace{-0.9cm}\Omega^1=\lambda\frac{\partial}{\partial t}+\left[\mu_1\frac{\partial X}{\partial x}+\mu_2\frac{\partial Y}{\partial x}+(\dot{\mu}_1-\dot{x}\dot{\lambda}_1)\frac{n}{x}\right]\frac{\partial}{\partial X}\nn\\
&&\hspace{4cm}+\left[\mu_1\frac{\partial X}{\partial y}+\mu_2\frac{\partial Y}{\partial y}+(\dot{\mu}_2-\dot{y}\dot{\lambda})\frac{n}{y}\right]\frac{\partial}{\partial Y}\equiv\Xi.
\eea

As the symmetry vector fields $\Sigma$ and $\Xi$ are for the same equation and therefore the infinitesimal symmetries must also be equal.  Comparing the above two equations we get the following relations,
\bea
\xi=\lambda,\quad \frac{n}{x}\dot{\mu}_1+\mu_1\frac{\partial X}{\partial x}+\mu_2\frac{\partial Y}{\partial x}=\frac{\dot{\eta}_1}{U}-\eta_1\frac{\dot{U}}{U^2}-f(x,y)\dot{\xi},\label{coupsymeq1}\\
\frac{n}{y}\dot{\mu}_2+\mu_1\frac{\partial X}{\partial y}+\mu_2\frac{\partial Y}{\partial y}=\frac{\dot{\eta}_2}{V}-\eta_2\frac{\dot{V}}{V^2}-g(x,y)\dot{\xi}.\label{coupsymeq2}
\eea
We note here that the above equations are relations connecting the known point symmetries of the linear ODEs to symmetries of the nonlinear ODEs.  Solving these coupled equations one can obtain the symmetries for the nonlocal equation.  However,  we find that the general solution of the above equation cannot be given for arbitrary forms of $f$ and $g$.  The forms of $f$ and $g$ have to be suitably chosen to decouple the above system of equations.
In order to decouple the equations (\ref{coupsymeq1}) and (\ref{coupsymeq2}) we consider the relation
\bea
\frac{U}{V}=\frac{x^n}{y^n}e^{\int (f-g) dt}.
\eea
For the specific choice $f=g$, the nonlocal part in  the above equation vanishes and we obtain
\bea
\frac{U}{V}=\frac{x^n}{y^n}=Z.
\eea
The symmetry vector in terms of the new variable $Z$ becomes
\bea
&&\Lambda=\xi\frac{\partial}{\partial t}+\frac{1}{V}\eta_1\frac{\partial}{\partial Z}-\frac{U}{V^2}\eta_2\frac{\partial}{\partial Z},\\
&&\Omega=\lambda\frac{\partial}{\partial t}+n\mu_1\frac{x^{n-1}}{y^n}\frac{\partial}{\partial Z}-n\mu_2\frac{x^n}{y^{n+1}}\frac{\partial}{\partial Z}.
\eea
Comparing the above two equations we get
\bea
\lambda=\xi,\quad\quad \mu_1=\frac{x}{n}\left(\frac{\eta_1}{U}-\frac{\eta_2}{V}\right)+\frac{x}{y}\mu_2.\label{mu1mu2relation}
\eea
Substituting this in the symmetry determining equation (\ref{coupsymeq2}) we get
\bea
\hspace{-1cm}\dot{\mu}_2+\frac{\mu_2}{n}\left((x+y)f_y-n\frac{\dot{y}}{y}\right)=xy\left(\frac{\eta_2}{V}-\frac{\eta_1}{U}\right)+\frac{y}{n}\frac{d}{dt}\left(\frac{\eta_2}{V}\right)-\frac{y}{n}f(x,y)\dot{\xi}.\label{decoupledeq}
\eea
Solving the above linear first order ODE with the substitution of following point symmetries of the linear system (\ref{2dlinear}) \cite{hydon,head:1993,ibragimov:1999}, 
\bea
&&\hspace{-2cm}\Lambda_1=\frac{\partial}{\partial t},\quad\Lambda_2=\frac{\partial}{\partial U},\quad\Lambda_3=\frac{\partial}{\partial V}\quad\Lambda_4=t\frac{\partial}{\partial t},\quad\Lambda_5=t\frac{\partial}{\partial U},\quad\Lambda_6=t\frac{\partial}{\partial V},\nonumber\\
&&\hspace{-2cm}\Lambda_7=U\frac{\partial}{\partial t},\quad\Lambda_8=V\frac{\partial}{\partial t},\quad\Lambda_9=U\frac{\partial}{\partial V},\quad\Lambda_{10}=V\frac{\partial}{\partial U},\quad\Lambda_{11}=U\frac{\partial}{\partial U}+V\frac{\partial}{\partial V},\nonumber\\
&&\hspace{-2cm}\Lambda_{12}=U\frac{\partial}{\partial U}-V\frac{\partial}{\partial V},\quad\Lambda_{13}=t^2\frac{\partial}{\partial t}+Ut\frac{\partial}{\partial U}+Vt\frac{\partial}{\partial V},\nonumber\\
&&\hspace{-2cm}\Lambda_{14}=Ut\frac{\partial}{\partial t}+U^2\frac{\partial}{\partial U}+UV\frac{\partial}{\partial V},\quad\Lambda_{15}=Vt\frac{\partial}{\partial t}+UV\frac{\partial}{\partial U}+V^2\frac{\partial}{\partial V},\label{lie-point}
\eea
one can deduce a set of nonlocal symmetries associated with the nonlinear ODE (\ref{gencoupledeq}).  

\hspace{14.5cm}$\square$
\vskip 6pt
We find that Eq. (\ref{decoupledeq}) is a first order linear ODE whose solution can be deduced straightforwardly and therefore we consider a simple case and obtain the corresponding nonlocal symmetry.  For this purpose we consider the symmetry vector $\Lambda_{11}$ in Eq. (\ref{lie-point}). Substituting this in Eq. (\ref{decoupledeq}) and solving the resultant equation we get
\bea
\mu_2=ye^{-\frac{1}{n}\int(x+y)f_ydt},\qquad f_y=\frac{\partial f}{\partial y}.
\eea
Substituting this in (\ref{mu1mu2relation}) we find $\mu_1$ and the symmetry vector field corresponding to $\Lambda_{11}$ is given as
\bea
\Omega_{11}=xe^{-\frac{1}{n}\int(x+y)f_ydt}\frac{\partial}{\partial x}+ye^{-\frac{1}{n}\int(x+y)f_ydt}\frac{\partial}{\partial y}.
\eea

\n{\bf Proposition 5:} \emph{The nonlocal symmetry $\Omega_{11}$ reduces Eq. (\ref{gencoupledeq}), with $f(x,y)=g(x,y)$, to the integrable Riccati equations $\frac{dz_1}{dt}=-nz_1^2$, and $ \frac{dz_2}{dt}=-nz_2^2$ through the reduction transformations $z_1=\frac{\dot{x}}{x}+\frac{1}{n}f$, and $z_2=\frac{\dot{y}}{y}+\frac{1}{n}f$. }

\n\emph{Proof :} The Lagrange's system associated with the symmetry vector  $\Omega_{11}$  is
\bea
\frac{dt}{0}=\frac{dx}{x}=\frac{dy}{y}=\frac{d\dot{x}}{\dot{x}-\frac{x}{n}(x+y)f_y}=\frac{d\dot{y}}{\dot{y}-\frac{y}{n}(x+y)f_y}
\eea
The characteristics of this system are $t$, $z_1=\frac{\dot{x}}{x}+\frac{1}{n}f$, and $z_2=\frac{\dot{y}}{y}+\frac{1}{n}f$, and the reduced equations become
\bea
\frac{dz_1}{dt}=-nz_1^2,\qquad \frac{dz_2}{dt}=-nz_2^2.
\eea
The solution of the above system is
\bea
z_1=\frac{1}{I_1+nt},\qquad z_2=\frac{1}{I_2+nt},
\eea
where $I_1$ and $I_2$ are integration constants.  \hspace{8cm}$\square$

Substituting these in the expressions in the reduction transformations, and rearranging we get,
we get
\bea
\dot{x}=\frac{x}{I_1+nt}-\frac{x}{n}f,\qquad\dot{y}=\frac{y}{I_2+nt}-\frac{y}{n}f.\label{cmee-first-order}
\eea
We note that the above set of first order coupled ODEs is integrable only for specific forms of $f(x,y)$.  For the choice $f(x,y)=g(x,y)=a_1x+a_2y$, Eq. (\ref{gencoupledeq}) reduces to the following system of coupled modified Emden type equation \cite{gladwin:jpa:2009,gladwin:jmp:2010},
\begin{eqnarray}
&&\ddot{x}+2(a_1x+a_2y)\dot{x}+(a_1\dot{x}+a_2\dot{y})x+
(a_1x+a_2y)^2x=0,\nonumber\\
&&\ddot{y}+2(a_1x+a_2y)\dot{y}+(a_1\dot{x}+a_2\dot{y})y+
(a_1x+a_2y)^2y=0.\label{secondordercmee}
\end{eqnarray}
 By solving the corresponding system of first order ODEs (\ref{cmee-first-order}), the general solution of (\ref{secondordercmee}) can be obtained as
\bea
&&x(t)=\frac{2I_1(I_2+t)}{a_1I_1(2I_4+(2I_2+t)t)+a_2(2I_4+(2I_3+t)t)},\nonumber\\
&&y(t)=\frac{(I_3+t)}{a_1I_1(2I_4+(2I_2+t)t)+a_2(2I_4+(2I_3+t)t)},
\eea
where $I_3$ and $I_4$ are two more integration constants, and the general solution agrees with the known result \cite{gladwin:jpa:2009,gladwin:jmp:2010}.
\section{Conclusion}
\label{conclusion}
In this paper, we have developed a new systematic procedure to deduce the nonlocal symmetries of a class of arbitrary order nonlinear ODEs.  The procedure uses the knowledge of the Lie point symmetries of the linear equations and the nonlocal transformation connecting the linear and the nonlinear ODEs.  We note here that the order of the linear and the corresponding nonlinear equation remains the same.  
The procedure is illustrated for the second and third order ODEs with examples and the procedure is shown to be applicable to arbitrary order equations as well.  Using these nonlocal symmetries we  have constructed the general solution certain specific nonlinear ODEs. 
We also find that an $m^{th}$ order ODE of the form (\ref{arbitrarynonlinear}) with arbitrary $f(x)$ can be reduced to an $(m-1)^{th}$ order equation of the Riccati chain.
Further, we have extended the procedure to second order coupled  ODEs and obtained the general solution of the coupled modified Emden equation using the associated nonlocal symmetries. 
\section*{Acknowledgments}
The work forms a part of a research project of MS and an IRHPA
project and Ramanna Fellowship project of ML sponsored by the Department of Science \& Technology
(DST), Government of India.  ML is also supported by a Department of Atomic Energy Raja Ramanna Fellowship.

%\appendix
\section*{Appendix}
{\bf Demonstration of the correctness of nonlocal symmetries}

%\Appendix{jkljdf}
In this section we briefly illustrate that the nonlocal symmetries obtained using the procedure discussed in Sec. \ref{procedure} indeed satisfies the invariant condition (\ref{invariance}).  In order to do so, we consider as a specific example the following nonlocal symmetry vector (\ref{eg-appendix}) of Eq. (\ref{2gen2}),
\bea
&&\hspace{-1cm}\Omega_2=\left(\frac{x^{-n}}{n}e^{\int(\frac{x}{n}f_x- f)dt}
-\frac{1}{n^2}
\int x^{1-n}f_xe^{\int(\frac{1}{n}xf_x-f)dt}dt\right)
xe^{-\frac{1}{n}\int x f_xdt}
\frac{\partial}{\partial x}.\qquad\nonumber%(\ref{eg-appendix})
\eea
The symmetry invariance condition is given as
\bea
\left(\lambda\frac{\partial}{\partial t}+\mu\frac{\partial}{\partial x}+\mu^{(1)}\frac{\partial}{\partial \dot{x}}+\mu^{(2)}\frac{\partial}{\partial \ddot{x}}\right)(\ddot{x}-\phi(x,\dot{x}))=0,\label{invariance2}
\eea
where $\phi(x,\dot{x})=-((n-1)\frac{\dot{x}^2}{x}+2\dot{x}f+\frac{1}{n}x\dot{x}f_x+\frac{x}{n}f^2)$, the first prolongation $\mu^{(1)}=\dot{\mu}-\dot{x}\dot{\lambda}$ and the second prolongation $\mu^{(2)}=\frac{d}{dt}\left(\mu^{(1)}\right)-\ddot{x}\dot{\lambda}$.  From the symmetry vector (\ref{eg-appendix}) we find 
\bea
\hspace{-2cm}\lambda=0, \qquad \mu=\left(\frac{x^{-n}}{n}e^{\int( \frac{x}{n}f_x- f)dt}\label{mu-appendix}
-\frac{1}{n^2}
\int x^{1-n}f_xe^{\int(\frac{1}{n}xf_x-f)dt}dt\right)
xe^{-\frac{1}{n}\int x f_xdt}.
\eea
 Therefore we find that $\mu^{(1)}=\frac{d\mu}{dt}$ and $\mu^{(2)}=\frac{d\mu^{(1)}}{dt}$.  Substituting these in the symmetry invariance condition we find
\bea
-\mu\phi_x-\mu^{(1)}\phi_{\dot{x}}+\mu^{(2)}=0. \label{invariance3}
\eea
Differentiating $\mu$ with respect to $t$, we find $\mu^{(1)}$ and $\mu^{(2)}$.  Substituting these in the above equation we find $\mu$ given by  Eq. (\ref{mu-appendix}) satisfies the symmetry invariant condition (\ref{invariance3}).  Similarly one finds that all the other remaining nonlocal symmetries of Eq. (\ref{2gen2}) satisfy the symmetry invariant condition.

We wish to note that  the general form of the nonlocal symmetries for the class of ODEs (\ref{gen-arbitrary}) of an arbitrary finite order $m$ is obtained by solving (\ref{arbitrary-sym}) and is given as
\bea
\mu=\mbox{e}^{p}\left[C+\int \frac{g_x}{g}\mbox{e}^{-p}\left[\frac{d}{dt}\left(\frac{\eta }{x}e^{-\int fdt}\right)-\xi f_t-f\dot{\xi}-\frac{d}{dt}\left(\frac{\xi g_t}{g}\right)\right]dt\right],\label{gen-nonlocal-sym}
\eea
where $C$ is an integration constant, $p=-\int\frac{g}{g_x}\left(f_x+\frac{\partial}{\partial x}\left(\frac{\dot{g}}{g}\right)\right)dt$,  $\eta$ and $\xi$ are the point symmetries of the linear ODE (\ref{gen-arbitrarylinear}).  One can verify that the above deduced general form of nonlocal symmetry satisfies the symmetry invariance condition (\ref{invariance}) for an arbitrary finite order $m$ as in the case of $\Omega_2$ above.  It is also straightforward to check that the specific forms of $\mu$ used in finding the generators $\Omega_i$ for the various examples in Secs. \ref{procedure} - \ref{coupled-symmetry} follow from (\ref{gen-nonlocal-sym}).
\section*{References}
%\bibliography{sym}
%\end{document}
\providecommand{\newblock}{}

\end{document}